\documentclass[twocolumn,prb,showpacs,citeautoscript,floatfix]{revtex4}
\usepackage{array}
\usepackage{multirow}
\usepackage{amssymb}
\usepackage{graphicx}
\input{epsf}

\begin{document}

\title{Polarization-dependent infrared reflectivity study of Sr$_{2.5}$Ca$_{11.5}$Cu$_{24}$O$_{41}$ under pressure: Charge dynamics, charge distribution, and anisotropy}
\author{S. Frank$^{1}$, A. Huber$^{1}$, U. Ammerahl$^{2}$, M. Huecker$^{3}$, and C.~A.~Kuntscher$^{1,*}$}
\affiliation{$^{1}$Experimentalphysik~2,~Universit\"at~Augsburg,~
D-86195~Augsburg,~Germany }
\affiliation{$^{2}$ Laboratoire de Physico-Chimie de L'{\'E}tat Solide, ICMMO, UMR 8182, Universit{\'e} Paris-Sud, 91405
Orsay, Cedex, France}
\affiliation{$^{3}$ Condensed Matter Physics and Materials Science Department, Brookhaven National Laboratory, Upton, New York 11973, USA}

\date{\today}

\begin{abstract}
We present a polarization-dependent infrared reflectivity study of the spin-ladder compound Sr$_{2.5}$Ca$_{11.5}$Cu$_{24}$O$_{41}$ under pressure.
The optical response is strongly anisotropic, with the highest reflectivity along the ladders/chains (\textbf{E}$\|$c) revealing a metallic character. For the polarization direction perpendicular to the ladder plane, an insulating behavior is observed. With increasing pressure the optical conductivity for \textbf{E}$\|$c shows a strong increase, which is most pronounced below 2000~cm$^{-1}$.
According to the spectral weight analysis of the \textbf{E}$\|$c optical conductivity the hole concentration in the ladders increases with increasing pressure and tends to saturate at high pressure. At $\sim$7.5~GPa the number of holes per Cu atom in the ladders has increased by $\Delta \delta$=0.09 ($\pm$0.01), and the Cu valence in the ladders has reached the value +2.33. The optical data suggest that Sr$_{2.5}$Ca$_{11.5}$Cu$_{24}$O$_{41}$ remains electronically highly anisotropic up to high pressure, also at low temperatures.
\end{abstract}


\maketitle

\section{Introduction}

The quasi-one-dimensional spin ladder compounds Sr$_{14-x}$Ca$_{x}$Cu$_{24}$O$_{41}$ have been studied extensively due to the emergence of superconductivity for high Ca content and high pressure \cite{Dagotto96,Vuletic06,Kim06,Dagotto92,Uehara96,Abbamonte04}. The theoretically \cite{Dagotto96} predicted superconducting state was first observed in Sr$_{0.4}$Ca$_{13.6}$Cu$_{24}$O$_{41.8}$ below T$_c$=12 K and pressures
$\geq$3~GPa \cite{Uehara96}. The crystal structure of Sr$_{14-x}$Ca$_{x}$Cu$_{24}$O$_{41}$ consists of two types of copper oxide layers that are parallel to the crystallographic a-c plane and alternate along the b axis: the Cu$_2$O$_3$ planes which contain the two-leg ladders, and CuO$_2$ planes containing chains with edge-shared CuO$_4$ plaquettes \cite{Vuletic06}.
At ambient conditions the parent compound Sr$_{14}$Cu$_{24}$O$_{41}$ has an intrinsic hole doping of six holes per formula unit, which results in a average Cu valence of +2.25. Although the substitution of Sr by isovalent Ca does not change the intrinsic charge concentration in the system, the physical properties alter drastically, which has been attributed to a redistribution of hole carriers among the ladder and chain subsystems \cite{Vuletic06}.

Among the key issues for understanding the mechanism of superconductivity in Sr$_{14-x}$Ca$_{x}$Cu$_{24}$O$_{41}$ are the distribution of charge carriers among the ladders and chains, and the dimensionality of the system. Both aspects can be addressed by polarization-dependent infrared spectroscopy, which is also applied in the present study.
According to optical studies for the undoped parent compound Sr$_{14}$Cu$_{24}$O$_{41}$, at ambient conditions one hole resides on the ladders and five holes on the chains \cite{Osafune97}. These values are close to those found by XAS and NMR experiments \cite{Huang13,Nucker00, Piskunov05,comment1}. Ca doping changes the carrier distribution as was recently discussed and summarized in Ref.\ \onlinecite{Huang13}: Despite some discrepancies in the absolute number of holes, it is now generally accepted that Ca doping triggers a chemical pressure induced transfer of holes from the chains to the ladders.

Besides a high Ca content, high pressure is needed to induce superconductivity in Sr$_{14-x}$Ca$_{x}$Cu$_{24}$O$_{41}$ \cite{Uehara96,Motoyama02}. The pressure-dependent charge distribution in Sr$_{14-x}$Ca$_{x}$Cu$_{24}$O$_{41}$ has been studied by NMR measurements for $x$=0 and $x$=12 for pressures up to 3.2~GPa. According to this study the hole concentration on the ladders increases by $\Delta\delta$\,$\approx$\,0.03 per Cu atom in Sr$_{2}$Ca$_{12}$Cu$_{24}$O$_{41}$, when a pressure of 3.2~GPa is applied. Furthermore, pressure-dependent electrical transport measurements on Sr$_{2.5}$Ca$_{11.5}$Cu$_{24}$O$_{41}$ suggest that the superconducting state has a quasi-two-dimensional character.\cite{Nagata98}

In the present study we investigate the charge distribution in Sr$_{2.5}$Ca$_{11.5}$Cu$_{24}$O$_{41}$ up to a high pressure of 7.5~GPa by infrared spectroscopy. Additionally, we address the dimensionality of the system under pressure by presenting polarization-dependent infrared spectra at room temperature and at low temperature.

\begin{figure*}
\includegraphics[width=14cm]{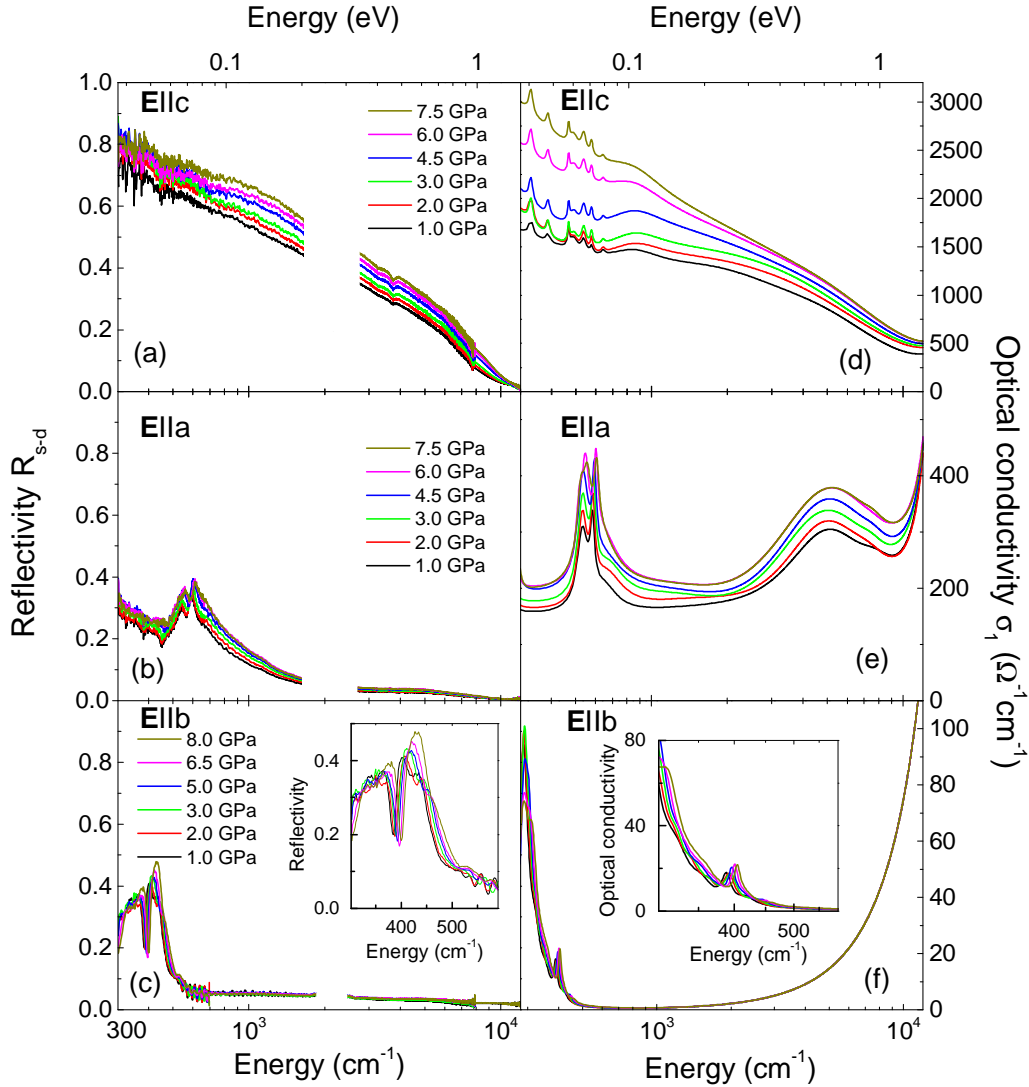}
\caption{(a)-(c): Reflectivity spectra R$_{s-d}$ of Sr$_{2.5}$Ca$_{11.5}$Cu$_{24}$O$_{41}$ for the polarization {\bf E} of the radiation along the three crystal axes at room temperature. (d)-(f): Real part of the optical conductivity $\sigma_1$ obtained from the Drude-Lorentz fitting of the reflectivity spectra.}
\label{fig:reflcond}
\end{figure*}

\section{Experiment}
The studied Sr$_{2.5}$Ca$_{11.5}$Cu$_{24}$O$_{41}$ single crystal was grown by the traveling-solvent floating-zone \cite{Ammerahl99}.
The polarization-dependent room-temperature reflectivity as a function of pressure was measured over a broad frequency range (300 - 12000~cm$^{-1}$) using a Bruker IFs 66v/S Fourier transform infrared spectrometer.
Two diamond-anvil pressure cells were employed for pressure generation:
A clamp diamond-anvil cell (Diacell cryoDAC-Mega)
and a Syassen-Holzapfel diamond-anvil cell (DAC) \cite{Keller77}.
Finely ground CsI powder served as quasi-hydrostatic pressure transmission medium
to ensure direct contact of the sample with the diamond anvil. The sample size was 80 x 80 $\mu$m$^2$ for the high-frequency range and about 200 x 200 $\mu$m$^2$ for frequencies below 700~cm$^{-1}$ to avoid diffraction effects.
To focus the infrared beam onto the small sample in the pressure cell, a Bruker IR Scope II infrared microscope with a 15$\times$ magnification objective was used.
Spectra taken at the inner diamond-air interface of the empty cell served as the
reference for normalization of the sample spectra (see Ref.\ \onlinecite{Frank06} for an illustration of the measurement geometry). The absolute reflectivity at the sample-diamond interface, denoted as $R_{s-d}$, was calculated according to $R_{s-d}(\omega)=R_{\rm dia}\times I_{s}(\omega)/I_{d}(\omega)$, where $I_s(\omega)$
denotes the intensity spectrum reflected from the sample-diamond interface and $I_d(\omega)$ the reference
spectrum of the diamond-air interface.
For $R_{\rm dia}$ a value of 0.167 was calculated from the refractive index of diamond $n_{\rm dia}$, which is assumed to be independent of pressure.

The polarization-dependent reflectivity measurements for frequencies 780 - 6000~cm$^{-1}$ at low temperature and
high pressure were carried out using a home-built infrared
microscope coupled to the FTIR spectrometer and maintained
at the same vacuum conditions, in order to avoid absorption
lines of H$_2$O and CO$_2$ molecules. Details about the home-built infrared
microscope can be found in Ref. \onlinecite{Kuntscher14}. A Syassen-Holzapfel DAC \cite{Keller77} for the pressure generation was mounted in a continuous-flow helium
cryostat (Cryo Vac KONTI cryostat). More details about the geometry of the reflectivity measurements can be found in
our earlier publications.\cite{Pashkin06,Kuntscher06} As reference, we used the intensity
reflected from the silver coated steel gasket inside the DAC.
Correspondingly, the absolute reflectivity at the sample-diamond interface $R_{s-d}$ was calculated according to $R_{s-d}(\omega)=I_{s}(\omega)/I_{Ag}(\omega)$, where $I_s(\omega)$
denotes the intensity spectrum reflected from the sample-diamond interface and $I_{Ag}(\omega)$ the reference
spectrum of the diamond-silver interface.
All reflectivity spectra shown in this paper refer to the absolute reflectivity at
the sample-diamond interface R$_{s-d}$.
The pressure in the DAC was determined {\it in situ} by the standard ruby-fluorescence technique.\cite{Mao86}

\begin{figure}[t]
 \begin{center}
   \includegraphics[scale=0.3]{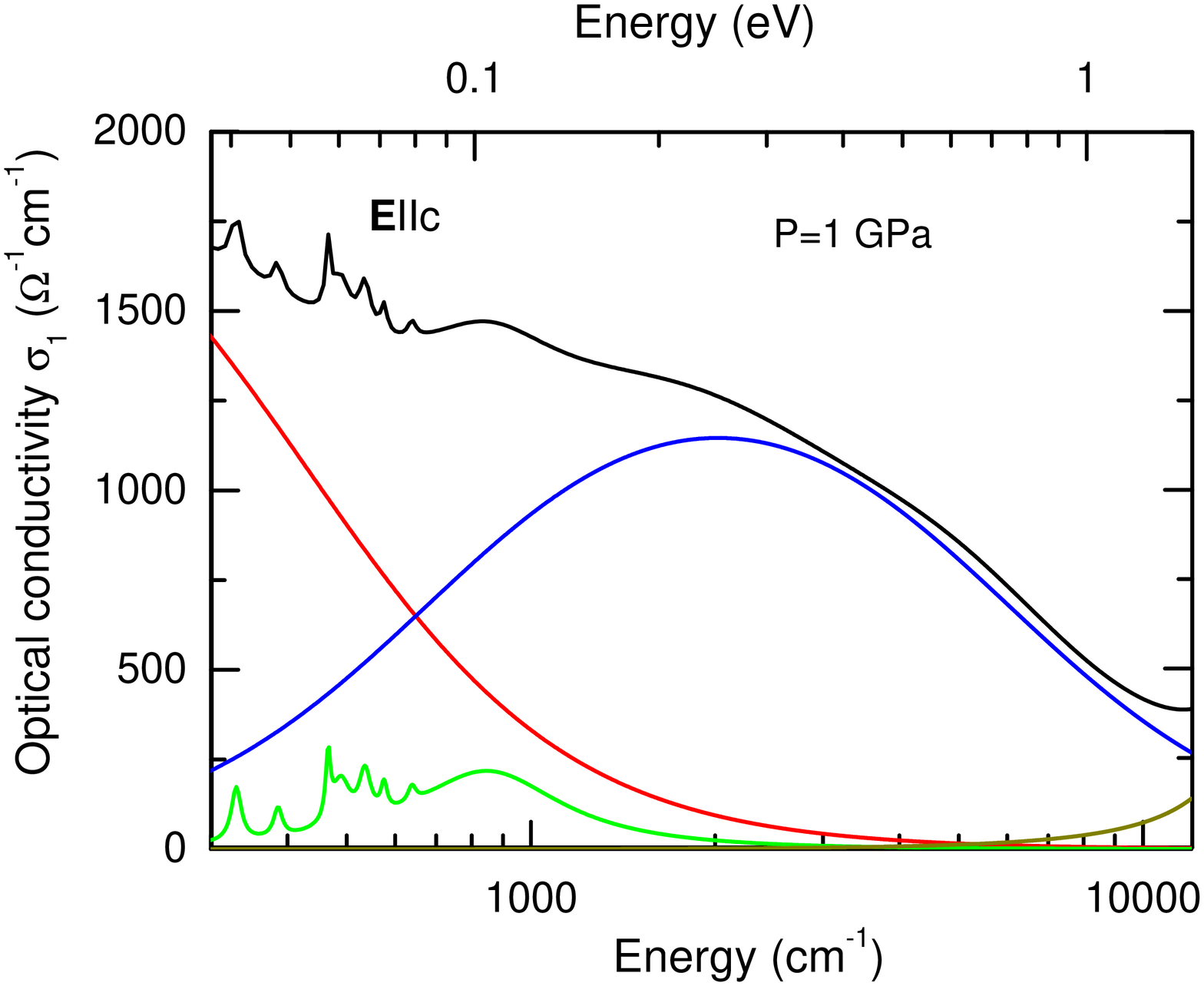}\\
   \caption{Real part of the optical conductivity $\sigma_1$ for \textbf{E}$\|$c at the lowest applied pressure at room temperature and the various contributions (Drude term, MIR band, phonon modes) as obtained from the Drude-Lorentz fits.}
   \label{fig:ambient}
  \end{center}
\end{figure}

\begin{figure}[t]
 \begin{center}
   \includegraphics[scale=0.3]{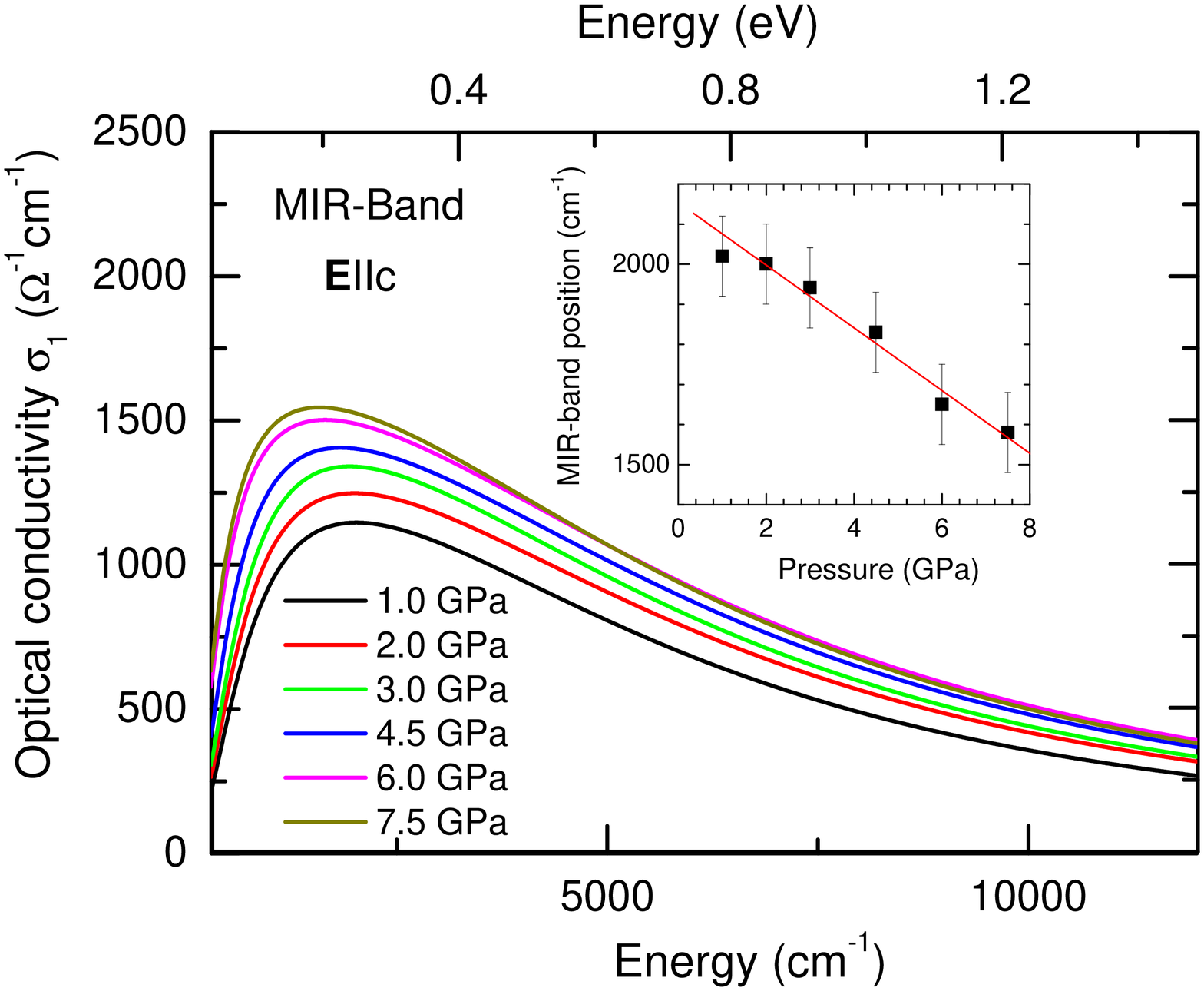}\\
   \caption{MIR band as a function of pressure at room temperature, as obtained from the Drude-Lorentz fits of the reflectivity spectra. Inset: Position of the MIR band as a function of pressure.}
   \label{fig:MIR}
  \end{center}
\end{figure}

\begin{figure*}
   \includegraphics[width=15cm]{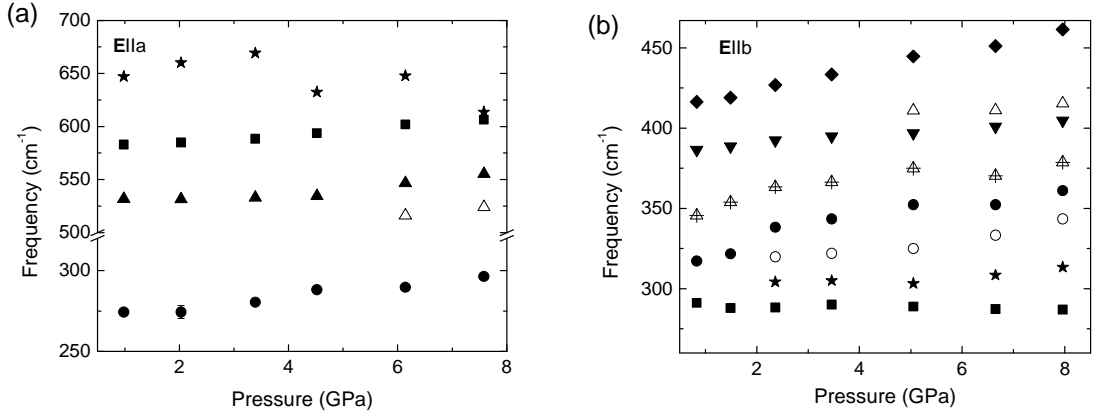}\\
   \caption{Frequencies of the phonon modes as a function of pressure at room temperature, as obtained from the Drude-Lorentz fits of the reflectivity spectra, for (a) \textbf{E}$\|$a and (b) \textbf{E}$\|$b.}
   \label{fig:phonon}
\end{figure*}

\begin{figure}
   \includegraphics[width=6.5cm]{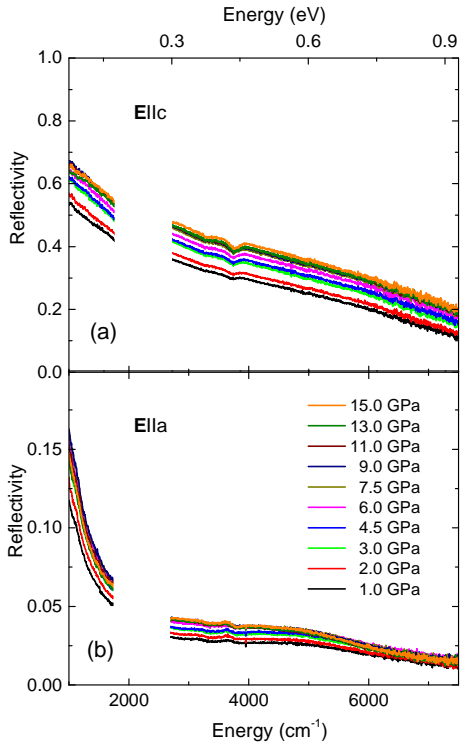}\\
   \caption{Reflectivity spectra of Sr$_{2.5}$Ca$_{11.5}$Cu$_{24}$O$_{41}$ for (a) \textbf{E}$\|$c and (b) \textbf{E}$\|$a up to 15~GPa at room temperature.}
   \label{fig:refl}
\end{figure}

\section{Results and Analysis}

The pressure-dependent reflectivity spectra of Sr$_{2.5}$Ca$_{11.5}$Cu$_{24}$O$_{41}$ at room temperature are depicted in Fig.~\ref{fig:reflcond}(a)-(c) for the polarization direction along the ladders/chains (\textbf{E}$\|$c), along the rungs (\textbf{E}$\|$a), and perpendicular to the ladder plane (\textbf{E}$\|$b), respectively. Features in the frequency range 1700 - 2700~cm$^{-1}$ are artifacts originating from
multiphonon absorptions in the diamond anvils, which are not fully corrected by the normalization procedure, and are not considered in the following analysis of the spectra.
The polarization-dependent reflectivity spectra at the lowest applied pressure clearly reveal the electronic anisotropy of Sr$_{2.5}$Ca$_{11.5}$Cu$_{24}$O$_{41}$, consistent with earlier results \cite{Osafune99,Ruzicka98,Osafune97}:
For \textbf{E}$\|$c, i.e., along the ladders and chains, the overall reflectivity is high, whereas for \textbf{E}$\|$a
it is considerably lower, but still reveals a metallic behavior (see also the corresponding optical conductivity as described below). For \textbf{E}$\|$b a typical insulating behavior is found, with an overall low reflectivity and strong phonon excitations in the low-frequency range.
As compared to the results from ambient-pressure measurements on Sr$_{2.5}$Ca$_{11.5}$Cu$_{24}$O$_{41}$ the observed phonon modes have a lower intensity and are broadened.
According to the pressure-dependent reflectivity data the electronic anistropy of the sample is preserved up to the highest applied pressure ($\approx$8~GPa): For \textbf{E}$\|$c the reflectivity increases monotonically with increasing pressure. Also for \textbf{E}$\|$a a gradual increase is found, which is however much lower as compared to the \textbf{E}$\|$c direction. Along the insulating \textbf{E}$\|$b direction the reflectivity does not change with pressure, except the alterations related to the phonon mode excitations.

The real part of the optical conductivity $\sigma_1$ was obtained by fitting the
reflectivity spectra with the Drude-Lorentz model\cite{Wooten72} combined with the
normal-incidence Fresnel equation
\begin{equation}
R_{s-d} =\left| \frac{n_{\rm dia}-\sqrt{\epsilon_s}}{n_{\rm
dia}+\sqrt{\epsilon_s}}\right|^2 , \epsilon_s = \epsilon_\infty +
\frac{i \sigma}{\epsilon_0 \omega} \quad ,
\end{equation}
where $n_{dia}$ is the refractive index of diamond and $\epsilon_s$ the complex dielectric function of the
sample. $\epsilon_{\infty}$ is the optical dielectric constant.
The so-obtained optical conductivity spectra are shown in Fig.~\ref{fig:reflcond}(d)-(f) for the three polarization directions. The optical conductivity is highest for \textbf{E}$\|$c and the metallic character is clearly revealed by a Drude contribution. The various contributions to the optical conductivity are illustrated in Fig.~\ref{fig:ambient}: Besides the Drude term a pronounced mid-infrared absorption band (MIR band) centered at around 2000~cm$^{-1}$ and phonon excitations are found.
For \textbf{E}$\|$a the conductivity at the lowest frequencies is finite indicating a metallic character. At around 5000~cm~$^{-1}$ an absorption band is observed followed by the onset of higher-frequency excitations. For \textbf{E}$\|$b the low-energy optical conductivity contains phonon excitations besides higher-frequency excitations with the main spectral weight above the measured frequency range.

With increasing pressure a strong increase in the \textbf{E}$\|$c optical conductivity is observed, which is most pronounced below 2000~cm$^{-1}$ [see Fig.~\ref{fig:reflcond}(d)]. Based on the Drude-Lorentz fit of the pressure-dependent reflectivity spectra, the MIR band was extracted and is depicted in Fig.\ \ref{fig:MIR}. From the maxima we have estimated the energy position of the MIR band, which we plot in the inset of Fig. 3 as a function of pressure.
One can see that with increasing pressure the MIR band shows a red shift. Simultaneously, its spectral weight increases.
For the polarization \textbf{E}$\|$a the overall optical conductivity increases with increasing pressure
[see Fig.\ \ref{fig:reflcond}(e)].
The frequencies of the phonon modes were obtained from the Drude-Lorentz fits and are depicted in Fig.\ \ref{fig:phonon}(a). The phonon modes show a slight hardening with increasing pressure. The mode at 530~cm$^{-1}$ splits into two modes above 4.5~GPa.
For \textbf{E}$\|$b the overall optical conductivity is pressure-independent, except for the low-frequency range, where the phonon modes are observed [see Fig.\ \ref{fig:reflcond}(f) and its inset]. The pressure dependence of the phonon frequencies from the Drude-Lorentz fits are plotted in Fig.\ \ref{fig:phonon}(b). The phonon modes harden with increasing pressure. Above 2~GPa additional, but weak phonon modes appear, and above 4.5~GPa the phonon mode close to 400~cm$^{-1}$ splits.
In general, a pressure-induced symmetry change of the crystal structure will modify the phonon spectrum - with typical signatures being anomalies in the pressure-induced frequency shifts, mode intensity or width, or splitting of modes.
In the case of Sr$_{2.5}$Ca$_{11.5}$Cu$_{24}$O$_{41}$ the new high pressure modes appear as shoulders of the strong modes present at ambient conditions. These additional modes might already be present at ambient conditions and their intensity just increases upon pressure application due to charge redistribution, like transfer of charges from the chains to the ladders; therefore, we hesitate to interpret them in terms of pressure-induced crystal symmetry changes.
Also the non-hydrostatic components of the pressure in the DAC might play a role.
Furthermore, it is difficult to attribute the observed phonon modes to Cu--O vibrations of either chains or ladders, since the bond lengths are similar. In conclusion, the pressure dependence of the infrared-active phonon modes support the findings of x-ray diffraction measurements, which did not find indications for a pressure-induced structural phase transition up to 9~GPa \cite{Isobe98,Pachot99}.

Additional information on the pressure dependence of the electronic properties of Sr$_{2.5}$Ca$_{11.5}$Cu$_{24}$O$_{41}$ has been obtained by reflectivity measurements within the ladder plane for higher pressures, i.e., up to 15~GPa [see Fig.\ \ref{fig:refl}].
Along both \textbf{E}$\|$c and \textbf{E}$\|$a directions the reflectivity spectra barely change for pressures above $\sim$9~GPa. All above-described changes with pressure are reversible upon pressure release.

\section{Discussion}
\subsection{Pressure-induced hole transfer onto the ladders}

The high conductivity in Sr$_{14-x}$Ca$_{x}$Cu$_{24}$O$_{41}$ along the $c$ direction has been attributed to the charge carriers of the ladder subunits \cite{Motoyama97,Mueller98,Osafune97}. Optical studies of Sr$_{14}$Cu$_{24}$O$_{41}$ at ambient conditions found an intrinsic hole doping of six holes per formula unit, with one hole in the Cu$_2$O$_3$ ladders and five holes in the CuO$_2$ chains at ambient conditions \cite{Osafune97}. This distribution of holes among ladders and chains is close to the one found by XAS and NMR experiments \cite{Huang13,Nucker00, Piskunov05}.
With increasing Ca doping the total carrier concentration in the material is conserved, but the physical properties change drastically, which was attributed to a redistribution of charge carriers from the chains to the ladders with increasing Ca content \cite{Motoyama97,Mueller98,Osafune97,Vuletic06,Deng11,Carter96,Rusydi07,Kabasawa08,Ilakovac12,Tafra08,Magishi98,
Piskunov05}.
Because the ionic radius of Ca is smaller than that of Sr, the lattice parameter $b$ decreases for increasing Ca content \cite{McCarron88}. Since a similar effect occurs when external pressure is applied,\cite{Isobe98,Pachot99} the substitution of Sr by Ca can be interpreted in terms of a chemical pressure. Thus one may ask, whether the application of external pressure causes a similar charge carrier redistribution.

The real part of the optical conductivity of Sr$_{2.5}$Ca$_{11.5}$Cu$_{24}$O$_{41}$ for \textbf{E}$\|$c (see Fig.\ \ref{fig:ambient}) consists of a Drude term and an MIR band. Upon pressure application an increase in the optical conductivity is observed, which is most pronounced for the frequency range below about 2000 cm$^{-1}$.
Obviously, this increase suggests a redistribution of spectral weight from high to low frequencies.
According to the sum rule (see below) the spectral weight is a measure of the effective charge carrier concentration N$_{eff}$.
The excitations below 1.2~eV ($\approx$\,9678~cm$^{-1}$) can be attributed to the ladders, whereas those above 1.2~eV
are related to the chains.\cite{Osafune97}
Therefore, the increase in the low-frequency optical conductivity is related to a transfer of holes from the chain to the ladder subsystem.

\begin{figure}[t]
 \begin{center}
   \includegraphics[scale=0.25]{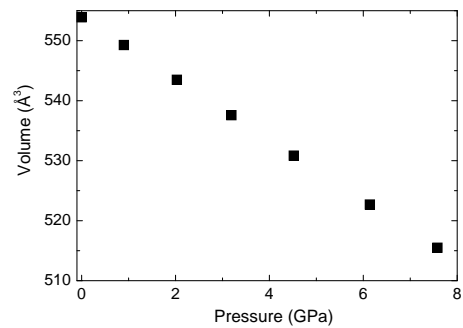}\\
   \caption{
    Estimated unit cell volume of Sr$_{2.5}$Ca$_{11.5}$Cu$_{24}$O$_{41}$ as a function of pressure at room temperature.}
   \label{fig:volume}
  \end{center}
\end{figure}

\begin{figure*}[t]
\includegraphics[scale=0.9]{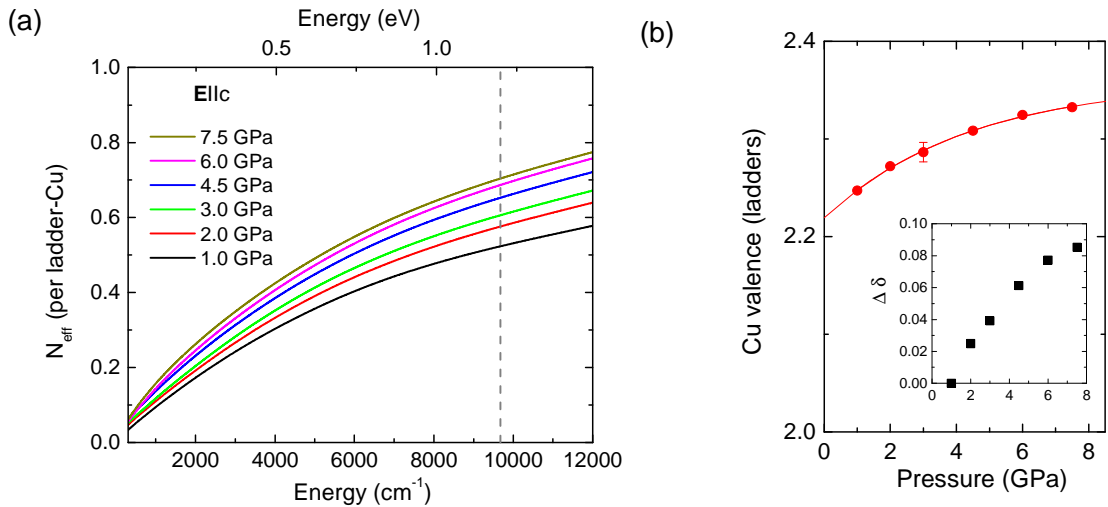}\\
\caption{(a) Effective charge concentration N$_{eff}$ per Cu atom obtained according to Equ.\ (\ref{eqn:lochkonz}).
(b) Cu valence in the ladders as a function of pressure. At high pressure the increase of holes in the ladders
saturates. Inset: Relative increase of the number of holes per Cu atoms in the ladders.}
\label{fig:Lochanzahl}
\end{figure*}

By applying the sum rule the number of charge carriers in the ladders can be calculated from the optical conductivity spectrum according to
\begin{equation}
N_{eff}(\omega)=\frac{2m^\ast V}{\pi e^2}
\int^{\omega_2}_{\omega_1} \sigma_1 (\omega^{'}) d\omega^{'} \quad ,
\label{eqn:lochkonz}
\end{equation}
where $m^\ast$ is the effective mass of the charge carriers, $V$ the pressure-dependent volume of the unit cell, and $\omega_1$=300~cm$^{-1}$, i.e., the lowest measured frequency.
Hereby, we assume that only the holes on the ladder subunits are mobile, i.e., the spectral weight for the conducting \textbf{E}$\|$c direction in the frequency range up to $\omega_2$=1.2~eV is related to the holes of the ladders \cite{Osafune97,comment2}.
Hence the calculated $N_{eff}$ gives the effective carrier concentration in the ladders per Cu atom $\delta$ according
to N$_{eff}$\,=\,A$\delta$, where $A$ is a constant determined according to literature data (see details given below).

To be able to calculate the effective charge carrier concentration $N_{eff}$ the pressure-dependent volume of the unit cell must be known. Since corresponding data are not available for the studied compound, we made use of the linear pressure coefficients for the lattice parameters of the closely related compound Sr$_{0.4}$Ca$_{13.6}$Cu$_{24}$O$_{41}$,\cite{Isobe98,Pachot99} for which values of $\alpha$\,=\,0,0093~\AA/GPa, $\beta$\,=\,0,0823~\AA/GPa, and $\gamma$\,=\,0,0074~\AA/GPa, respectively, can be found in the literature \cite{comment3}. These values were applied to the lattice parameters of Sr$_{2.5}$Ca$_{11.5}$Cu$_{24}$O$_{41}$. The so-obtained volume of the unit cell is depicted in Fig.\ \ref{fig:volume}.

To obtain the effective carrier concentration in the ladders per Cu atom, $\delta$, from the spectral weight according to N$_{eff}$\,=\,A$\delta$, the coefficient $A$ needs to be determined. This parameter was obtained based on published results for Sr$_{14}$Cu$_{24}$O$_{41}$, for which NEXAFS and optical measurements found one hole per formula unit on the ladders \cite{Nagata98,Osafune97}. With 14 Cu atoms on the ladder units per unit cell, this gives $\delta$\,=\,$\frac{1}{14}$\,=\,0.07 (corresponding ladder Cu Valence of +2.07). From the spectral weight analysis of the optical data of Osafune et al. \cite{Osafune97} we obtain N$_{eff}$\,$\approx$\,0.148, which gives A\,=$\frac{N_{eff}}{\delta}$\,=\,2.12.

For consistency check, the above-described procedure has been applied to our lowest-pressure data of
Sr$_{2.5}$Ca$_{11.5}$Cu$_{24}$O$_{41}$, which should deviate only slightly from the ambient-pressure data. We find V\,=\,549,26~{\AA$^3$} and the spectral weight 8.93$\cdot$10$^6$~$\Omega^{-1}$cm$^{-2}$ at $\omega_2$=1.2~eV, which gives N$_{eff}$($\omega_2$)\,=\,0.523 according to Equ.\ (\ref{eqn:lochkonz}). From this value the number of holes in the ladders per Cu atom is calculated according to $\delta$\,=$\frac{N_{eff}}{A}$\,=\,0.25$\pm$0.01, corresponding to a ladder Cu valence of +2.25.
This result is in agreement with Ca doping dependent infrared measurements \cite{Osafune97}, where for Sr$_{3}$Ca$_{11}$Cu$_{24}$O$_{41}$ a hole concentration of 0.2 holes per Cu atom on the ladders (corresponding ladder Cu valence: +2.20) was found, keeping in mind the slightly higher Ca content and the pressure offset of 1~GPa in our measurement.

The resulting effective charge concentration N$_{eff}$ per Cu atom and the Cu valence of the ladder subunits as a function of applied pressure are depicted in Fig.\ \ref{fig:Lochanzahl} (a) and (b). With increasing pressure the
Cu valence and the related number of holes in the ladders increases and tends to saturate at high pressure. At $P$$\sim$7.5~GPa the number of holes per Cu atom in the ladders has increased by 0.09 ($\pm$0.01) and the Cu valence in the ladders has reached the value +2.33 [see Fig.\ \ref{fig:Lochanzahl} (b)].
Hence, both Ca substitution and the application of external pressure lead to an increase of the charge carrier concentration on the ladders.
Pressure-dependent NMR measurements on Sr$_{2}$Ca$_{12}$Cu$_{24}$O$_{41}$ observed a pressure-induced increase of $\Delta\delta$\,$\approx$\,0.03 at $\sim$3.2~GPa \cite{Piskunov05}, which compares well with the value $\Delta \delta$\,=\,0.04$\pm$0.01 at around 3~GPa according to our IR measurements [see inset of Fig.\ \ref{fig:Lochanzahl} (b)].

Reflectivity spectra at higher pressures ($>$8~GPa) could only be obtained in the higher-frequency range (mid- and near-infrared) due to the diffraction limit and are plotted in Fig.\ \ref{fig:refl}. Obviously, the overall reflectivity increases with increasing pressure for both polarization directions \textbf{E}$\|$c and \textbf{E}$\|$a up to $\sim$9~GPa and are basically constant above this pressure. From this behavior we infer that the spectral weight along the conducting $c$ direction and the associated concentration of charge carriers on the ladders are approximately constant above $\sim$9~GPa.

\subsection{Nature of the MIR band}

The optical conductivity spectrum of Sr$_{2.5}$Ca$_{11.5}$Cu$_{24}$O$_{41}$ for the polarization \textbf{E}$\|$c consists of a pronounced MIR band, whose origin will be discussed in the following based on its pressure dependence.
Photoemission experiments at 130~K found an electronic band located at $\sim$0.5~eV (4032 cm$^{-1}$) for Sr$_{14}$Cu$_{24}$O$_{41}$, which shifts toward the Fermi energy with increasing Ca content \cite{Takahashi97,Sato98,Yokoya97}. The band was interpreted in terms of strong electronic correlations.
Within the single-band Hubbard model for correlated electron systems characteristic contributions are expected in the optical conductivity spectrum \cite{Rozenberg96,Rozenberg95,Imada98,Jarrell95}. For the metallic solution of the Hubbard model a Drude term and an excitation band in the mid-infrared frequency range are observed in the optical conductivity spectra, consistent with experimental findings \cite{Kezsmarki04,Kezsmarki06,Kuntscher06,Kuntscher08}.
These contributions are also present in the optical conductivity spectrum of Sr$_{2.5}$Ca$_{11.5}$Cu$_{24}$O$_{41}$ (see Fig.\ \ref{fig:ambient}). With increasing pressure the MIR band shifts to smaller frequencies, which is consistent with a pressure-induced bandwidth increase. Although the spectral weight of the MIR band is large compared to that of the Drude term \cite{Jarrell95}, an interpretation of the MIR band within the Hubbard model appears reasonable.

On the other hand, the importance of electron-phonon interaction in Sr$_{14-x}$Ca$_{x}$Cu$_{24}$O$_{41}$ has been discussed earlier \cite{Takahashi97,Kaneshita06}. In fact, MIR absorption features are fingerprints for the excitation of polaronic quasiparticles. An interpretation of the MIR band in Sr$_{14-x}$Ca$_{x}$Cu$_{24}$O$_{41}$ in terms of the dissociation of bipolarons composed of holes on one rung was suggested in Refs. \cite{Osafune97,Hayward96}.
Generally, polaronic excitations cause a characteristic MIR band in the optical conductivity spectrum, whose energy position is related to the polaron binding energy and which is expected to shift to lower frequencies with increasing pressure \cite{Emin93,Kuntscher05,Thirunavukkuarasu06,Frank06,Ebad-Allah13}.
The pressure-induced shift of the MIR band in Sr$_{2.5}$Ca$_{11.5}$Cu$_{24}$O$_{41}$ to lower frequencies is thus consistent with the polaron picture.

Furthermore, it has been shown theoretically that in Sr$_{14-x}$Ca$_{x}$Cu$_{24}$O$_{41}$ a pressure-induced phase transition from single polarons to bipolarons can occur \cite{Kaneshita06}. During the formation of bipolarons two polarons are localized within the same potential well \cite{Emin93}. The absorption band of small bipolarons generally appears at higher energies compared to single small polarons, with the frequency position $ \omega_p\,=\,4E_b-U $, where E$_b$ is the polaron binding energy and $U$ the Coulomb interaction between the two polarons. Hence, a pressure-induced phase transition of single polarons to bipolarons would result in a shift of the MIR band to higher frequencies. This is not observed in our data, and therefore such a phase transition seems to be unlikely.

\begin{figure}[t]
\includegraphics[scale=0.30]{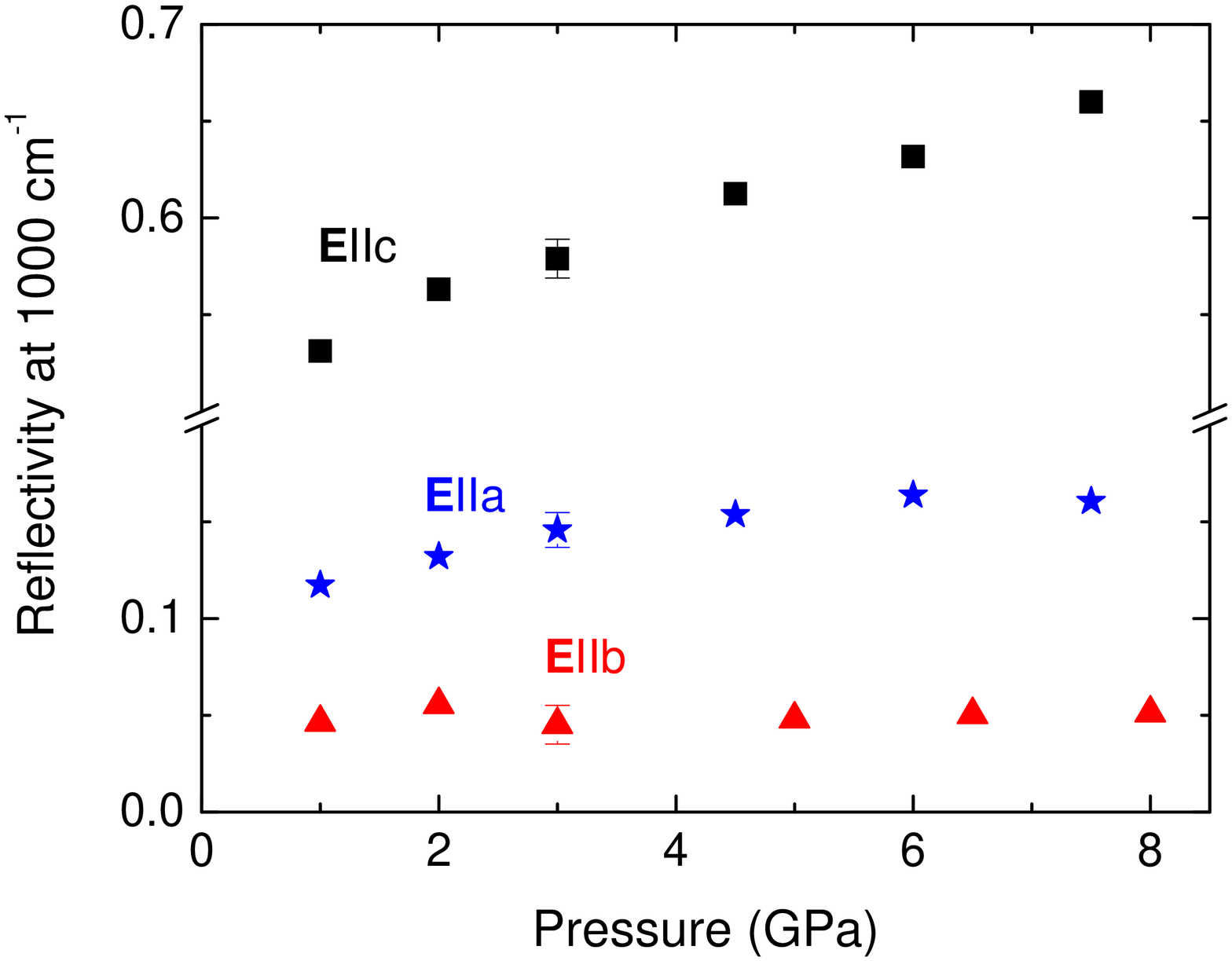}\\
\caption{Room-Temperature reflectivity of Sr$_{2.5}$Ca$_{11.5}$Cu$_{24}$O$_{41}$ at 1000~cm$^{-1}$ for \textbf{E}$\|$c, \textbf{E}$\|$a, and \textbf{E}$\|$b as a function of pressure.}
\label{fig:refllevel}
\end{figure}

\subsection{Pressure-induced dimensional crossover?}

In Sr$_{2.5}$Ca$_{11.5}$Cu$_{24}$O$_{41}$ superconductivity is observed at 4~GPa and 6~K \cite{Nagata98}.
Owing to the similarity in crystal structure with the high-temperature copper-oxide superconductors the question at issue is, whether the superconductivity in Sr$_{14-x}$Ca$_{x}$Cu$_{24}$O$_{41}$ is a one-dimensional or -- like in the high-temperature superconductors -- a two-dimensional phenomenon. It was suggested earlier that the superconductivity in the spin ladder compounds is of two-dimensional nature.\cite{Vuletic03,Nagata98,Piskunov05}
Pressure-dependent infrared data for various polarization directions can give insight into the anisotropy of a material.

According to Figs.\ \ref{fig:reflcond}(a) - (c) only for the direction \textbf{E}$\|$c the reflectivity -- and concomitant the optical conductivity -- significantly increases under pressure. For illustration we plot in Fig.\ \ref{fig:refllevel} the reflectivity level along the three polarization directions at 1000~cm$^{-1}$, i.e., outside the phonon mode region, which is representative for the overall behavior of the reflectivity spectra.
For \textbf{E}$\|$b the reflectivity is unchanged under pressure, and for \textbf{E}$\|$a only a small increase of the reflectivity with increasing pressure is observed. Hence, no strong increase of the reflectivity level for the directions perpendicular to \textbf{E}$\|$c is observed. Therefore, a pressure-induced dimensional crossover towards a more two-dimensional character at room temperature seems to be unlikely according to our data.

\begin{figure}[t]
\includegraphics[scale=0.3]{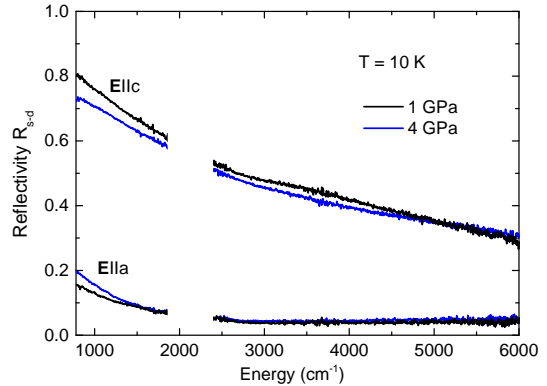}\\
\caption{Reflectivity spectra R$_{s-d}$ of Sr$_{2.5}$Ca$_{11.5}$Cu$_{24}$O$_{41}$ at 10~K for the lowest ($\sim$1~GPa) and highest ($\sim$4~GPa) pressure for the two polarization directions \textbf{E}$\|$c and \textbf{E}$\|$a within the ladder/chain plane.}
\label{fig:low-T}
\end{figure}

An electrical transport study suggested that pressurized Sr$_{2.5}$Ca$_{11.5}$Cu$_{24}$O$_{41}$ becomes two-dimensional at low temperature \cite{Nagata98}.
To test the dimensionality of the system within the ladder plane at low temperature, we carried out additional reflectivity measurements at various pressures. Fig.\ \ref{fig:low-T} depicts the reflectivity spectra at 10~K for the polarization direction along the two crystal axes within the ladder/chain plane, i.e., along the $c$ and $a$ axis, at the lowest ($\sim$1~GPa) and highest ($\sim$4~GPa) applied pressure. Obviously, the anisotropy within the ladder/chain plane at base temperature is only slightly affected by pressure, and a crossover to two-dimensional electronic properties in Sr$_{2.5}$Ca$_{11.5}$Cu$_{24}$O$_{41}$ is not observed up to 4~GPa within the studied frequency range.

\section{Conclusions}

In conclusion, our polarization-dependent infrared reflectivity study shows that  Sr$_{2.5}$Ca$_{11.5}$Cu$_{24}$O$_{41}$ is electronically highly anisotropic with a conducting behavior along the ladders and an insulating behavior perpendicular to the ladder plane. Along the in-plane direction perpendicular to ladders the metallic character is lower as compared to the ladder direction.
The pressure-induced increase in the reflectivity and the corresponding optical conductivity is strongest for the polarization direction along the ladders, \textbf{E}$\|$c.
The pronounced MIR band in the \textbf{E}$\|$c optical conductivity spectrum red shifts under pressure, which can be attributed to strong electronic correlations or to the excitation of polaronic quasiparticles.
The hole concentration in the ladders increases with increasing pressure and tends to saturate at high pressure, resulting in a Cu valence in the ladders of +2.33 at $P$$\sim$7.5~GPa, which corresponds to an increase in the number of holes per ladder Cu atom by 0.09 ($\pm$0.01).
Above 9~GPa the MIR reflectivity within the ladder plane is independent of pressure.
Sr$_{2.5}$Ca$_{11.5}$Cu$_{24}$O$_{41}$ remains highly anisotropic up to high pressure and at low temperatures within the studied frequency range.

\section{Acknowledgment}

This work is financially supported by the DFG (KU1432/6-1). Work at Brookhaven was supported by the Office of Basic Energy
Sciences (BES), Division of Materials Science and Engineering, U.S. Department of Energy
(DOE), under Contract No. DE-AC02-98CH10886.

{}


\begin{thebibliography}{}
\item[$^*$] Email: christine.kuntscher@physik.uni-augsburg.de

\bibitem{Dagotto96}
E. Dagotto and T. M. Rice, Science {\bf 271}, 618 (1996).

\bibitem{Vuletic06}
T. Vuletic, B. Korin-Hamzic, T. Ivek, S. Tomic, B. Gorshunov, M. Dressel, and J. Akimitsu,
Phys.\ Rep.\ {\bf 428}, 169 (2006).

\bibitem{Kim06}
B. J. Kim, H. Koh, E. Rotenberg, S.-J. Oh, H. Eisaki, N.
Motoyama, S. Uchida, T. Tohyama, S. Maekawa, Z.-X. Shen
et al., Nat. Phys. {\bf 2}, 397 (2006).

\bibitem{Dagotto92}
E. Dagotto, J. Rieira, and D. Scalapino, Phys. Rev. B {\bf 45}, 5744 (1992).

\bibitem{Uehara96}
M. Uehara, T. Nagata, J. Akimitsu, H. Takahashi1, N. Mori, and K. Kinoshita,
J. Phys.\ Soc.\ Jpn.\ {\bf 65}, 2764 (1996).

\bibitem{Abbamonte04}
P. Abbamonte, G. Blumberg, A. Rusydi, A. Gozar, P. G. Evans, T. Siegrist, L. Venema,
H. Eisaki, E. D. Isaacs, and G. A. Sawatzky,
Nature (London) {\bf 431}, 1078 (2004).

\bibitem{Osafune97}
T. Osafune, N. Motoyama, H. Eisaki, and S. Uchida, Phys.\ Rev.\ Lett.\ {\bf 78}, 1980 (1997).

\bibitem{Nucker00}
N. N\"ucker, M. Merz, C. A. Kuntscher, S. Gerhold, S. Schuppler, R. Neudert, M. S. Golden, J. Fink,
D. Schild, S. Stadler, V. Chakarian, J. Freeland, Y. U. Idzerda, K. Conder, M. Uehara, T. Nagata,
J. Goto, and J. Akimitsu, N. Motoyama, H. Eisaki, S. Uchida, U. Ammerahl, and A. Revcolevschi,
Phys.\ Rev.\ B {\bf 62}, 14384 (2000).

\bibitem{Piskunov05}
Y. Piskunov, D. Jerome, P. Auban-Senzier, P. Wzietek, and A. Yakubovsky,
Phys.\ Rev.\ B {\bf 72}, 064512 (2005).

\bibitem{Huang13}
M.-J. Huang, G. Denk. Y. Y. Chin, Z. Hu, J.-G. Cheng, F. C. Chou, K. Conder, J.-S. Zhou, T.-W. Pi,
J. B. Goodenough, H.-J. Lin, and C. T. Chen, Phys.\ Rev.\ B {\bf 88}, 014520 (2013).

\bibitem{comment1}
Please note that the values obtained by the recent XAS study by Rusydi et al. \cite{Rusydi07} differ largely from other XAS results.

\bibitem{Motoyama02}
N. Motoyama, H. Eisaki, S. Uchida, N. Takeshita, N. Mori, T. Nakanishi, and H. Takahashi, Europhys. Lett. {\bf 58}, 758 (2002).

\bibitem{Nagata98}
T. Nagata, M. Uehara, J. Goto, J. Akimitsu, N. Motoyama, H. Eisaki, S. Uchida, H. Takahashi, T. Nakanishi, and N. Mori, Phys. Rev. Lett. {\bf 81}, 1090 (1998).

\bibitem{Ammerahl99}
U. Ammerahl and A. Revcolevschi, J. Cryst. Growth {\bf 197}, 825 (1999).

\bibitem{Keller77}
R. Keller and W. B. Holzapfel, Rev. Sci. Instrum. {\bf 48}, 517
(1977); G. Huber, K. Syassen and W. B. Holzapfel, Phys.
Rev. B {\bf 15}, 5123 (1977).

\bibitem{Frank06}
S. Frank, C. A. Kuntscher, I. Loa, K. Syassen, and F. Lichtenberg, Phys. Rev. B {\bf 74}, 054105 (2006).

\bibitem{Kuntscher14}
C. A. Kuntscher, A. Huber, and M. H\"ucker, Phys. Rev. B {\bf 89}, 134510 (2014).

\bibitem{Pashkin06}
A. Pashkin, M. Dressel, and C. A. Kuntscher, Phys. Rev. B {\bf 74}, 165118 (2006).

\bibitem{Kuntscher06}
C. A. Kuntscher, S. Frank, A. Pashkin, M. Hoinkis,
M. Klemm, M. Sing, S. Horn, and R. Claessen, Phys. Rev. B {\bf 74}, 184402 (2006).

\bibitem{Mao86}
H. K. Mao, J. Xu, and P. M. Bell, J. Geophys. Res. {\bf 91},
4673 (1986).

\bibitem{Osafune99}
T. Osafune, N. Motoyama, H. Eisaki, S. Uchida, and S. Tajima, Phys. Rev. Lett.\ {\bf 82}, 1313 (1999).

\bibitem{Ruzicka98}
B. Ruzicka, L. Degiorgi, U. Ammerahl, G. Dhalenne, and A. Revcolevschi,
Eur.\ Phys.\ J. B {\bf 6}, 301 (1998).

\bibitem{Wooten72}
F. Wooten, Optical Properties of Solids (Academic, New York, 1972).

\bibitem{Isobe98}
M. Isobe, T. Ohta, M. Onoda, F. Izumi, S. Nakano, J. Y. Li, Y. Matsui, E. Takyama-Muromachi, T. Matsumoto, and H. Hayakawa, Phys. Rev. B {\bf 57}, 613 (1998).

\bibitem{Pachot99}
S. Pachot, P. Bordet, R. Cava, C. Chaillout, C. Darie, M. Hanfland, M. Marezio,
and H. Takagi, Phys. Rev. B {\bf 59}, 12048 (1999).

\bibitem{Motoyama97}
N. Motoyama, T. Osafune, T. Kakeshita, H. Eisake, and S. Uchida, Phys. Rev. B {\bf 55}, R3386 (1997).

\bibitem{Mueller98}
T. F. A. M\"uller, V. Anisimov, T. M. Rice, I. Dasgupta, and T. Saha-Dasgupta,
Phys. Rev. B {\bf 57}, R12655 (1998).

\bibitem{Carter96}
S. A. Carter, B. Batlogg, R. J. Cava, J. J. Krajewski, W. F. Peck, Jr., and T. M. Rice,
Phys. Rev. Lett. {\bf 77}, 1378 (1996).

\bibitem{Tafra08}
E. Tafra, B. Korin-Hamzic, M. Basletic, A. Hamzic, M. Dressel, and J. Akimitsu,
Phys. Rev. B {\bf 78}, 155122 (2008).

\bibitem{Magishi98}
K. Magishi, S. Matsumoto, Y. Kitaoka, K. Ishida, K. Asayama, M. Uehara, T. Nagata, and J. Akimitsu,
Phys. Rev. B {\bf 57}, 11533 (1998).

\bibitem{Rusydi07}
A. Rusydi, M. Berciu, P. Abbamonte, S. Smadici, H. Eisaki, Y. Fujimaki, S. Uchida, M. R\"ubhausen,
and G. A. Sawatzky,
Phys.\ Rev.\ B {\bf 75}, 104510 (2007).

\bibitem{Kabasawa08}
E. Kabasawa, J. Nakamura, N. Yamada, K. Kuroki, H. Yamazaki, M. Watanabe, J. D. Denlinger, S. Shin, and
R. C. C. Perera, J. Phys. Soc. Jpn. {\bf 77}, 034704 (2008).

\bibitem{Deng11}
G. Deng, V. Pomjakushin, V. Petricek, E. Pomjakushina, M. Kenzelmann, and K. Conder,
Phys.\ Rev.\ B {\bf 84}, 144111 (2011).

\bibitem{Ilakovac12}
V. Ilakovac, C. Gougoussis, M. Calandra, N. B. Brookes, V. Bisogni, S. G. Chiuzbajan, J. Akimitsu,
O. Milat, S. Tomic, and C. F. Hague, Phys.\ Rev.\ B {\bf 85}, 075108 (2012).

\bibitem{McCarron88}
E. M. McCarron, M. A. Subramanian, J. C. Calabrese, and R. L. Harlow, Mat. Res. Bull. {\bf 23}, 1355 (1988).

\bibitem{comment2}
The excitations of the holes in the chains are observed for frequencies larger than $\sim$2.5~eV \cite{Osafune97}, which is well above the measured frequency range.

\bibitem{comment3}
The linear pressure coefficient $\alpha$ ($\beta$, $\gamma$) was obtained from the pressure dependence of the lattice parameter $a$ ($b$, $c$) by a linear fit to the experimental data according to the equation
$a$($P$)=$a$($0$) + $\alpha$$P$, where $a$($P$) is the lattice parameter $a$ at a pressure $P$.



\bibitem{Takahashi97}
T. Takahashi, T. Yokoya, A. Ashihara, O. Akaki, H. Fujisawa, A. Chainani, M. Uehara,
T. Nagata, J. Akimitsu, and H. Tsunetsugu, Phys. Rev. B {\bf 56}, 7870 (1997).

\bibitem{Sato98}
T. Sato, T. Yokoya, T. Takahashi, M. Uehara, T. Nagata, J. Goto, and J. Akimitsu, J. Phys. Chem. Solids {\bf 59}, 1912 (1998).

\bibitem{Yokoya97}
T. Yokoya, H.-D. Kim, A. Ashihara, H. Kumigashira, H. Fujisawa, T. Takahashi,
M. Uehara, T. Nagata, and J. Akimitsu, Physica C {\bf 282}, 997 (1997).

\bibitem{Hayward96}
C. A. Hayward, D. Poilblanc, and D. J. Scalapino, Phys. Rev. B {\bf 53}, R8863 (1996).


\bibitem{Rozenberg96}
M. J. Rozenberg, G. Kotliar, and H. H. Kajueter, Phys. Rev. B {\bf 54}, 8452 (1996).

\bibitem{Rozenberg95}
M. J. Rozenberg, G. Kotliar, H. Kajueter, G. A. Thomas, D. H. Rapkine, J. M. Honig, and P. Metcalf,
Phys. Rev. Lett. {\bf 75}, 105 (1995).

\bibitem{Imada98}
M. Imada, A. Fujimori, and Y. Tokura, Rev. Mod. Phys. {\bf 70}, 1039 (1998).

\bibitem{Jarrell95}
M. Jarrell, J. K. Freericks, and T. Pruschke, Phys. Rev. B {\bf 51}, 11 704 (1995).

\bibitem{Kezsmarki04}
I. Kezsmarki, N. Hanasaki, D. Hashimoto, S. Iguchi, Y. Taguchi, S. Miyasaka, and Y. Tokura, Phys. Rev. Lett. {\bf 93}, 266401 (2004).

\bibitem{Kezsmarki06}
I. Kezsmarki, N. Hanasaki, K. Watanabe, S. Iguchi, Y. Taguchi, S. Miyasaka, and Y. Tokura, Phys. Rev. B {\bf 73}, 125122 (2006).

\bibitem{Kuntscher08}
C. A. Kuntscher, A. Pashkin, H. Hoffmann, S. Frank, M. Klemm, S. Horn, A. Schönleber, S. van Smaalen, M. Hanfland, S. Glawion, M. Sing und R. Claessen, Phys. Rev. B {\bf 78}, 035106 (2008).

\bibitem{Kaneshita06}
E. Kaneshita, I. Martin, and A. R. Bishop, Phys. Rev. B {\bf 73}, 094514 (2006).

\bibitem{Emin93}
D. Emin, Phys. Rev. B {\bf 48}, 13691 (1993).

\bibitem{Kuntscher05}
C.A. Kuntscher, S.Frank, I. Loa, K. Syassen, T.Yamauchi und Y.Ueda, Phys. Rev. B {\bf 71}, 220502(R) (2005).

\bibitem{Thirunavukkuarasu06}
K. Thirunavukkuarasu, F. Lichtenberg, and C. A. Kuntscher, J. Phys.: Condens. Matter {\bf 18}, 9173 (2006).

\bibitem{Ebad-Allah13}
J. Ebad-Allah, L. Baldassarre, M. Sing, R. Claessen, V. A. M. Brabers, and C. A. Kuntscher,
J. Phys.: Condens. Matter {\bf 25}, 035602 (2013).

\bibitem{Vuletic03}
T. Vuletic, B. Korin-Hamzic, S. Tomic, B. Gorshunov, P. Haas, T. Room, M. Dressel, J. Akimitsu,
T. Sasaki, and T. Nagata
Phys.\ Rev.\ Lett.\ {\bf 90}, 257002 (2003).

\end{thebibliography}
\end{document}